\documentclass[a4paper]{article}

\usepackage{INTERSPEECH2021}
\usepackage{multirow}

\title{Neural Speaker Embeddings for Ultrasound-based Silent Speech Interfaces}
\name{Amin Honarmandi Shandiz$^1$, L\'aszl\'o T\'oth$^1$, G\'abor Gosztolya$^2$, Alexandra
Mark\'o$^{3,4}$, \\Tam\'as G\'abor Csap\'o$^{4,5}$}
\address{
  $^1$Institute of Informatics, University of Szeged, Hungary\\
  $^2$MTA-SZTE Research Group on Artificial Intelligence, Szeged, Hungary \\
  $^3$Department of Applied Linguistics and Phonetics, E\"otv\"os Lor\'and University, Budapest, Hungary \\
  $^4$MTA-ELTE Lend\"ulet Lingual Articulation Research Group, Budapest, Hungary \\
  $^5$Department of Telecommunications and Media Informatics, \\
Budapest University of Technology and Economics, Budapest, Hungary }
\email{\{shandiz, tothl, ggabor\}@inf.u-szeged.hu,
marko.alexandra@btk.elte.hu, csapot@tmit.bme.hu }

\begin{document}

\maketitle
\begin{abstract}
Articulatory-to-acoustic mapping seeks to reconstruct speech from a recording of the articulatory movements, for example, an ultrasound video. Just like speech signals, these recordings represent not only the linguistic content, but are also highly specific to the actual speaker. Hence, due to the lack of multi-speaker data sets, researchers have so far concentrated on speaker-dependent modeling. Here, we present multi-speaker experiments using the recently published TaL80 corpus. To model speaker characteristics, we adjusted the x-vector framework popular in speech processing to operate with ultrasound tongue videos. Next, we performed speaker recognition experiments using 50 speakers from the corpus. Then, we created speaker embedding vectors and evaluated them on the remaining speakers. Finally, we examined how the embedding vector influences the accuracy of our ultrasound-to-speech conversion network in a multi-speaker scenario. In the experiments we attained speaker recognition error rates below 3\%, and we also found that the embedding vectors generalize nicely to unseen speakers. Our first attempt to apply them in a multi-speaker silent speech framework brought about a marginal reduction in the error rate of the spectral estimation step. 

\end{abstract}
\noindent\textbf{Index Terms}: silent speech interface, articulatory-to-acoustic mapping, speaker embedding, x-vector

\section{Introduction}


Silent Speech Interfaces (SSI) aim to convert silent (mouthed) articulation to audible speech~\cite{Denby2010,Schultz2017a,Gonzalez-Lopez2020}. Such SSI systems could aid the communication of the speaking impaired (e.g.\ patients after laryngectomy). The theoretical background is provided by articulatory-to-acoustic mapping (AAM), where articulatory data is recorded while the subject is speaking, and machine learning methods (typically deep neural networks (DNNs)) are applied to predict the speech signal from the articulatory input. The set of articulatory acquisition devices includes ultrasound tongue imaging (UTI)~\cite{Csapo2017c,Csapo2020c,Ribeiro2021,shandiz2021improving,shandiz2021voice}, Magnetic Resonance Imaging (MRI)~\cite{yu2021reconstructing}, electromagnetic articulography (EMA)~\cite{Wang2012a,Kim2017a,Taguchi2018}, permanent magnetic articulography (PMA)~\cite{Fagan2008,Gonzalez2017a,Gonzalez-Lopez2021}, surface electromyography (sEMG)~\cite{Maier-Hein2005,Janke2012,Wand2018}, electro-optical stomatography (EOS)~\cite{Stone2020a}, lip videos~\cite{Ephrat2017,Michelsanti2020}, or a multimodal combination of the above~\cite{Freitas2014}.



Although there are lots of studies on generating intelligible speech from the above biosignals, most of these were conducted on relatively small databases of just a single or a small number of speakers~\cite{Denby2010,Schultz2017a,Gonzalez-Lopez2020}. Meanwhile, all of the articulatory tracking devices are obviously highly sensitive to the actual speaker's anatomy. A further source of variance may come from the possible misalignment of the recording equipment. For example, for ultrasound recordings, the probe fixing headset has to be remounted onto the speaker for each recording session. This inevitably causes the recorded ultrasound videos to become slightly misaligned between each recording session. 

Although the properly working cross-session and cross-speaker methodologies are still missing, there have already been studies in this direction. Kim et al. investigated speaker-independent Silent Speech Recognition (SSR) using EMA with 12 healthy and laryngectomized speakers~\cite{Kim2017a}.  For EMG-based recognition, several signal normalization and model adaptation methods were investigated by Maier-Hein et al.~\cite{Maier-Hein2005}. Janke et al. studied session-independent sEMG over 16 sessions of a speaker, and the results showed that sEMG is quite robust to minor changes in the electrode placement~\cite{Janke2012}. Wand et al.\ utilized domain-adversarial DNN training to increase the session-independency of their EMG-based speech recognizer~\cite{Wand2018}. For EOS-based SSR with a small vocabulary, Stone and Birkholz found the speaker-independent average word accuracy to be relatively stable, varying between 56--62\%~\cite{Stone2020a}.

Ultrasound-based SSI systems, however, might be less robust, as slight changes in probe positioning causes shifts and rotations in the resulting image. To this end,  we examined on the session dependency of UTI-based direct speech synthesis, and we proposed a simple session adaptation method~\cite{Gosztolya2020}. Ribeiro et al. experimented with the classification of UTI images of phonetic segments~\cite{Ribeiro2019}, and found that speaker-dependent systems perform much better than speaker-independent ones, but adding speaker information in a simple form (e.g. mean ultrasound image) helps the model generalize to unseen speakers. 
The same authors reported that unsupervised model adaptation can improve the results for silent speech (but not for modal speech)~\cite{Ribeiro2021}. They also performed multi-speaker recognition and synthesis experiments where they applied $x$-vectors for speaker conditioning -- but they extracted the $x$-vectors from the acoustic data and not from the ultrasound~\cite{ribeiro2020tal}.

\subsection{Speaker embedding vectors}
Just like the above-mentioned biosignals, speech signals also contain speaker-specific factors that hurt the cross-speaker performance of speech processing systems. Various DNN training and adaptation methods have been proposed as remedies, perhaps the simplest being is to use auxiliary input features that encapsulate information on speaker characteristics. The classic such representation was the $i$-vector~\cite{Senior2014}, but its role has recently been taken by the $x$-vector~\cite{Snyder-xv}. The $x$-vector can be created by training a DNN for speaker classification, and then the activation values of an upper hidden layer are used as the speaker embedding vector in other tasks. Our goal here is to adjust this $x$-vector scheme to ultrasound tongue videos. To this aim, we train a DNN for speaker classification, using 3D blocks of adjacent ultrasound frames as input. First, we report the speaker recognition accuracy of this network in~\ref{sec:results-xv}. Next, we evaluate the speaker embedding vector provided by this network on a separate set of speakers via very simple speaker recognition experiments in~\ref{sec:results-xvsi}. Finally, we attempt to apply the embedding vectors as auxiliary input for a second network which is trained to perform speech synthesis, more precisely, to estimate spectral vectors from the ultrasound frames (\ref{sec:results-embed}). Before the experiments, brief descriptions are given regarding our SSI framework, $x$-vectors and the experimental conditions in Sections ~\ref{sec:SSI}, ~\ref{sec:xvector} and
~\ref{sec:exp}, respectively.


\section{The SSI framework}
\label{sec:SSI}

\begin{figure}[!t]
\centering
\includegraphics[width=0.33\textwidth]{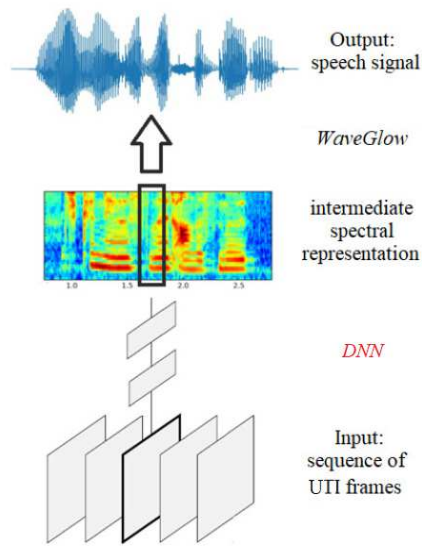}
\caption{\textit{Schematic diagram of the UTI-to-speech conversion process applied in our SSI model. 
}} \label{fig:framework}
\end{figure}

We presented our approach for creating an ultrasound-based SSI in~\cite{Csapo2020c}, so we just give a brief overview here. As Fig~\ref{fig:framework} shows, the input to our system is a sequence of ultrasound tongue imaging (UTI) frames, and the target sequence is a speech signal. This is a sequence-to-sequence mapping problem, which could be addressed by sophisticated encoder-decoder networks that would not even require aligned training data~\cite{ribeiro2020tal}. However, as we have synchronized input-output samples, most authors apply simpler networks that perform the mapping frame by frame~\cite{Tatulli2017, Csapo2020c}. The optimal output representation is also a matter of choice. While training DNNs that generate speech signals directly is feasible~\cite{waveglow}, it would require large amounts of training data. Alternatively, one can use a dense spectral representation as the training target that can be converted to speech~\cite{Csapo2017c}. Recently, speech synthesis technology has introduced neural vocoders for synthesizing speech from spectrograms~\cite{waveglow}. The main advantage of applying these in the SSI task is that that we can borrow large pre-trained networks for the speech synthesis step, and we have to deal only with the ultrasound-to-spectrum mapping task. In a recent study we used WaveGlow~\cite{Csapo2020c}, and we obtained higher quality speech than with standard vocoders~\cite{Csapo2017c}. 

As WaveGlow requires a mel-spectrogram with 80 spectral components as input, our DNN was trained to estimate such spectral vectors from the UTI video in a frame-by-frame manner. The input of our network is a 3D array of consecutive images, and the output is a 80-dimensional spectral vector. 
The convolutional (3D-CNN) network structure that we applied here~\cite{toth20203d} was the same as the lower, 'frame-level' part of the $x$-vector network, so we delay its presentation to the next section. As here the task is to estimate spectral vectors, we used a linear output layer and the network was trained to minimize the mean-squared error (MSE) of the regression task.

\section{Speaker embedding vectors for ultrasound tongue imaging}
\label{sec:xvector}

The $x$-vector concept was introduced by Snyder et al., motivated by the goal of replacing the previous Gaussian-based $i$-vector approach with a purely neural solution~\cite{Snyder-xv}. The basis behind the $x$-vector is a DNN that is trained to discriminate speakers. The network structure is unusual in the sense that~it consists of three main parts. The lower layers -- typically a time-delay network (TDNN~\cite{Peddinti2015}) -- operate on the level of frames. Then, the subsequent temporal pooling layer aggregates statistics over the frames of a given speech segment or utterance. This layer may collect only the mean values~\cite{LiMa}, the mean and the standard deviation~\cite{Snyder-xv, snyder-xv2, ShonTG18} or it may even apply a sophisticated attention mechanism~\cite{Okabe2018, Zhu2018selfatt}. The aggregated values are processed further by several fully connected layers. As these layer operate on the segment level, after training they can produce a fixed-size speaker embedding vector even from utterances of variable lengths. The embedding vector is typically obtained as the linear activation output of the fully connected layer right below or two layers below the softmax output~\cite{Snyder-xv}. 
\begin{figure}[!t]
\centering
\includegraphics[width=0.45\textwidth]{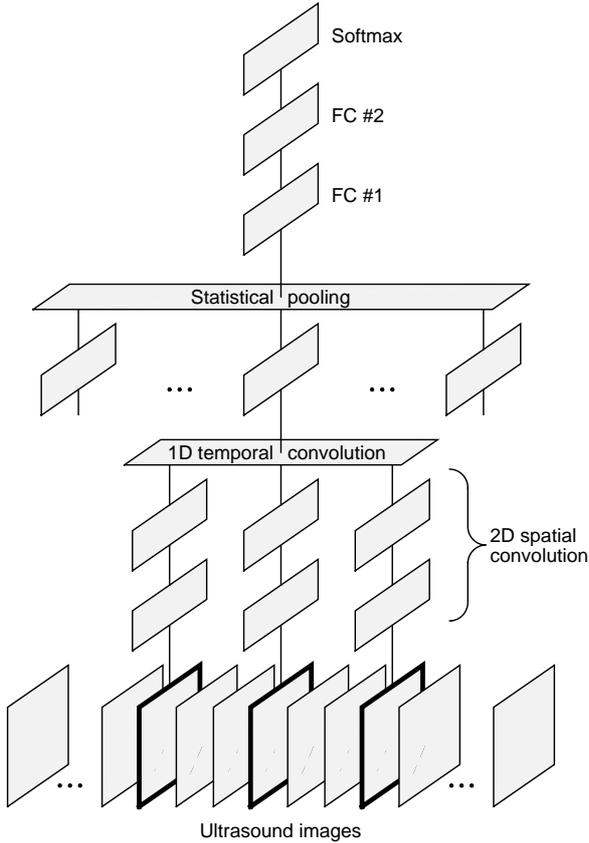}
\caption{\textit{Illustration of the UTI-based $x$-vector network.}} \label{fig:xvec}
\end{figure}

We adjusted the $x$-vector DNN to operate with a sequence of ultrasound images instead of a sequence of speech feature vectors as follows (see also Fig~\ref{fig:xvec}).

The {\bf frame-level part}  has the same structure as the network we apply in the spectral estimation step of Fig~\ref{fig:framework}~\cite{toth20203d}. Its input consists of 21 consecutive images, which are processed by a 3D convolutional layer in 5-image blocks, using a relatively large stride size of 4 along the time axis. The subsequent convolutional layers apply the 3D convolution in a decomposed ``(2+1)D'' form, first processing these blocks locally, and aggregating their content along time only at higher layers. This architecture was motivated by the findings of Tran et al. for video recognition~\cite{Tran}, and also by the success of TDNNs in speech recognition~\cite{Peddinti2015,Toth-conv2}. The top layer of the frame-level part is a dense layer that produces a local output vector at each frame position (ie., for the center frame of the input block).

The {\bf statistical pooling} layer performs a simple average pooling. We leave more complex solutions such as standard deviation pooling or attention-based weighting for future work.

The {\bf segment-level part} consists of two fully connected layers and a softmax output layer that has one output neuron for each speaker in the training set. The neural speaker embedding vector for a given input segment or utterance is extracted from one of these dense layers (see the experiments in Section~\ref{sec:results}).

\section{Experimental Set-Up}
\label{sec:exp}

In the experiments we used the TaL80 corpus~\cite{ribeiro2020tal}, which contains ultrasound, speech and lip video recordings from 81 speakers. Apart from the silent speech experiments, the speech signals were also recorded in parallel with the ultrasound, and here we used these synchronized ultrasound and speech tracks. The ultrasound was recorded using Articulate Instruments' Micro system that captures a midsaggital view of the tongue at a frame rate of 82 fps. The raw ultrasound images contain 64 x 842  data values, which were resized to 64 x 128 pixels. More details about the recording process can be found in~\cite{ribeiro2020tal}.

We divided the data into two sets, so we used 50 speakers to create the $x$-vector network, and we held out 31 speakers to train and evaluate the SSI network. Although the $x$-vector network is able to handle inputs with different sizes, for technical simplicity we chose to work with short uniform 2-second long chunks from each recording (similar to Shon et al.~\cite{ShonTG18}). To train the $x$-vector network, we extracted such 2-second (164 frames) chunks from each training file of the 50-speaker subset of the corpus, resulting in 76 chunks from each speaker on the average (minimum 65, maximum 85). Altogether we obtained 3800 such blocks, from which 2298 were used for training, 357 for development, and 1145 blocks for testing. 

The network used for the spectral estimation task in the SSI and the frame-level part of the $x$-vector network were structurally identical, consisting of 4 3D-convolutional layers (with a MaxPooling layer after every second convolution), and a fully connected layer that aggregates the convolutional outputs (the exact model parameters can be found in our earlier work~\cite{toth20203d}). The fully connected hidden layer was followed by a linear output layer for the spectral estimation task, while in the $x$-vector network the fully connected layer outputs were aggregated along time by average pooling. The segment-level part consisted of two fully connected layers (FC\#1 and FC\#2 in Fig~\ref{fig:xvec}) with sizes of 500 and 250, and the softmax output layer contained 50 neurons, corresponding to the 50 speakers of the train set. Both networks were trained using the Adam optimizer with a learning rate of 0.0002, using a batch size of 100 for the SSI network, and batch sizes of 4-16 for the $x$-vector network (using smaller batch sizes for longer segments). The nonlinearity applied in all hidden layers was the swish function.

\section{Results and Discussion}
\label{sec:results}

\begin{table}[t]
\caption{Speaker recognition error rates for the 50-speaker set as a function of the input segment duration.} \label{tab:xv}
\vspace{-2mm}
\centering
\renewcommand{\arraystretch}{1.1} 
\begin{tabular}{|l|c|c|}
\hline~~Segment length~ & ~Aggregated~ &  Error rate\\
~~(frames \& seconds) & frames & (Dev Set) \\
\hline
\hline
~~~21 (0.25 sec) & 1 & 3.16\%  \\
~~~41 (0.50 sec) & 21 & 2.87\%\\
~~~82~~(1.0 sec) & 62 & 2.10\%  \\
~~164 (2.0 sec) & 144 & 1.96\%  \\
\hline
\end{tabular}
\end{table}

\subsection{Training the $x$-vector network} 
\label{sec:results-xv}

In the first experiment we were interested to measure the speaker classification accuracy of our $x$-vector network, and how it is influenced by the input duration. The motivation for the aggregation step in the original $x$-vector model is that normally we have not just a single frame but whole utterances (at least a word) from a given speaker, and that certain phonetic segments may be less speaker-discriminative than others~\cite{ShonTG18} -- the most trivial example are the silent parts. But the situation is different when the input is an ultrasound tongue video, as the recording device always returns an image of the tongue, even when the speaker remains silent, and thus longer segments may not be necessary. In the first experiment we gradually increased the size of the input segment from 21 frames (when the lower part of the network returns one frame-level output, so effectively no temporal pooling occurs) to 164 frames (2 seconds).

Table~\ref{tab:xv} shows the speaker recognition error rates on the development set of the 50-speaker corpus for different durations of temporal pooling. As we expected, we obtained a very low error rate already with just a single-frame output, which shows that the UTI videos are very speaker-specific. Gradually increasing the segment length and thus the number of aggregated frames improves the results, but the improvements are smaller than those obtained, for example, in language recognition~\cite{SnyderLang}. 

\begin{figure}[!t]
\centering
\includegraphics[width=0.45\textwidth]{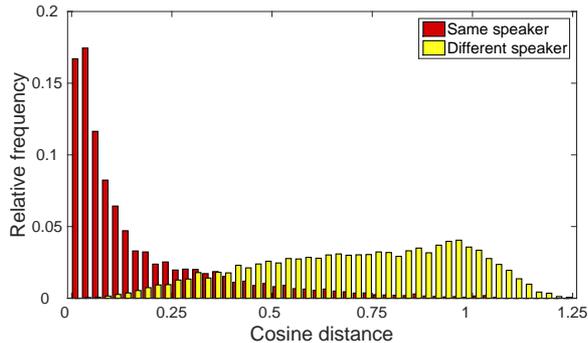}
\caption{\textit{Normalized histogram of the cosine distances for randomly chosen same-speaker and different-speaker $x$-vector pairs from the 31-speaker set.}} \label{fig:hist}
\end{figure}

\subsection{Validating the $x$-vectors on an independent speaker set} 
\label{sec:results-xvsi}

The second experiment sought to test the generalization ability of the $x$-vector. We were worried about overfitting the training speakers, as the popular $x$-vector implementations are trained with orders of magnitudes more training data and typically with more than a thousand speakers~\cite{Snyder-xv}. Thus, to validate our $x$-vector DNN, we performed a speaker recognition experiment on the dataset of the 31 speakers not used during $x$-vector training.

Using the speaker embeddings to recognize or verify a new set of speakers is not trivial, as the speakers will be different from those seen during training. For optimal performance, the embedding vectors are typically post-processed by factor analysis or dimension reduction methods~\cite{Snyder-xv,Dehak2009}. A simple and fast, though somewhat sub-optimal solution is to calculate the similarity score of two speakers based on the cosine similarity of their embedding vectors~\cite{Dehak2009}. According to this, we performed a simple speaker recognition experiment with a 1-nearest neighbor classifier using the cosine distance as follows. We fused the development and the test sets of the 31 speakers, and calculated the $x$-vectors for each segment. Then we performed leave-one-out classification, that is, to each vector in this set we found the closest one (excluding itself). The classification was considered correct if the speaker IDs of the two segments were identical. We intentionally used the simplest possible 1-NN classifier, as it is very sensitive to the accuracy of the distance function applied, and hence to the accuracy of the underlying $x$-vectors.

Snyder et al. defined the $x$-vector as the linear output of the first fully connected layer after statistical pooling~\cite{Snyder-xv}. However, some authors questioned whether this is the optimal strategy, and also whether dropping the nonlinearity is required~\cite{ShonTG18}. So we also examined which fully connected layer gives the best result, and if the nonlinearity is necessary or not.
The results are shown in Table~\ref{tab:1nn}. As can be seen, all embedding extraction methods resulted in very low speaker recognition error rates around 1-2\%. The best result was obtained when the speaker embedding was produced by the lower fully connected layer (FC\#2) without the nonlinearity. This coincides with the findings of Snyder et al.~\cite{Snyder-xv} and Shon et al.~\cite{ShonTG18}.

To further demonstrate the speaker discriminative abilities of the $x$-vectors on a new set of speakers, a histogram of the cosine distances is shown in Fig~\ref{fig:hist} for 10000 randomly selected same-speaker and different-speaker vector pairs from the 31-speaker dev+test subset. Although the distributions overlap slightly, this figure also suggests that the embedding vectors behave as expected, that is, vectors from the same speaker are closer to each other than vectors from different speakers.

\begin{table}[t]
\caption{Speaker recognition error rates for the held-out 31 speakers using 1-NN leave-one-out testing.} \label{tab:1nn}
\vspace{-2mm}
\centering
\renewcommand{\arraystretch}{1.1} 
\begin{tabular}{|l|c|c|c|c|}
\hline
~Embedding layer~ & \multicolumn{2}{c|}{\it FC\#1} & \multicolumn{2}{c|}{\it FC\#2} \\
\hline
Activation & swish & linear & swish & linear \\
\hline
Error rate & 0.96\% & 0.96\% & 2.03\% & 0.70\%\\
\hline
\end{tabular}
\vspace{5mm}

\caption{Mean squared error (MSE) for the SSI spectral estimation task in
the single-speaker and multi-speaker scenarios.} \label{tab:ssi}
\vspace{-2mm} \centering
\renewcommand{\arraystretch}{1.1}
\begin{tabular}{|l|c|c|c|}
\hline
~SSI Train+Test~ & ~Size of training ~& \multicolumn{2}{c|}{MSE}\\
\cline{3-4}
~~configuration~ & ~set (frames)~& Dev & Test\\
\hline\hline
single-speaker      & 254306 & 0.256 & 0.265 \\
\hline\hline
multi-speaker       & \multirow{2}{*}{305040} & 0.603 & 0.669\\
\cline{1-1} \cline{3-4}
multi-spk + Xvec    &   & 0.589 & 0.653\\
\hline
\end{tabular}
\end{table}

\subsection{Applying the speaker embedding in speech synthesis} 
\label{sec:results-embed}

In the last experiment we attempted to use the $x$-vector as an auxiliary input for the spectral estimator network of our SSI system (cf. Fig~\ref{fig:framework}). As the baseline, we trained the net with the single-speaker TaL1 corpus. The obtained mean squared error rates (MSE) shown in Fig~\ref{tab:ssi} are around 0.26, which results in a speech signal with a mel-cepstral distortion (MCD) of 3.12 after speech synthesis. This corresponds to a low-quality, but intelligible speech~\cite{Csapo2020c}. For multi-speaker training and testing we used samples from the 31-speaker subset. We did not use all the available data from the 31 speakers to demonstrate how the accuracy of the SSI network drops when switching to a multi-speaker scenario with a similar amount of training data. Table~\ref{tab:ssi} shows that the multi-speaker setup led to a drastically larger MSE. We note that the multi-speaker scenario is still speaker-dependent in the sense that the training, development and test sets are from the same speakers. Even larger performance drops can be expected in a speaker-independent configuration~\cite{Ribeiro2019}.

Finally, we re-trained the multi-speaker model with the $x$-vector as auxiliary input. However, the optimal way of combining the ultrasound input with the $x$-vector is not trivial. As the ultrasound images are processed by convolution, simple concatenation was not an option. We decided to inject the $x$-vector into the network only after the convolutional layers, which were initialized by transfer learning from the multi-speaker model. However, this solution might be suboptimal and requires further studies. We mention that Ribeiro et al. used the mean ultrasound frame to represent the speaker, so they could add this image to the CNN input as a second channel~\cite{Ribeiro2019}. 

The bottom row of Table~\ref{tab:ssi} shows that although the introduction of the $x$-vector resulted in a consistent improvement for both the development and the test sets, this improvement is marginal. Ribeiro et al. reported similarly small gains from using the speaker mean as the speaker-characteristic input~\cite{Ribeiro2019}. 

\section{Conclusions}
\label{sec:conlusion}

Here, we adjusted the $x$-vector framework of speech processing to ultrasound tongue videos to create a speaker-characteristic embedding vector. We modified the network architecture to process videos as input, and trained the network for speaker recognition. We obtained very low speaker recognition error rates, and our embedding vectors also seem to generalize well to new speakers. However, our first attempts to apply the embedding vector in a multi-speaker SSI scenario resulted in just a minimal improvement, showing that further studies are required on the proper application of the $x$-vector in this field.

\section{Acknowledgements}

This study was supported by grant NKFIH-1279-2/2020 of the Ministry for Innovation and Technology, Hungary, and by the Ministry of Innovation and the 
NRDI Office within the framework of the Artificial Intelligence National Laboratory Programme and also through project FK 124584. 
The GPU card used was donated by the NVIDIA Corporation. 

\clearpage

\bibliographystyle{IEEEtran}

\bibliography{Interspeech2021,ref_collection_csapot_nourl}

\end{document}